\documentclass[twocolumn,showpacs,preprintnumbers,
                aps,prd,amssymb,amsmath,a4paper]{revtex4}
\usepackage{latexsym}
\usepackage{bm}
\newcommand*{\zed}[1]{\mathbb{Z}_{{#1}}}

\newcommand*{\vect}[1]{\bm{{#1}}}

\newcommand*{\powerspectrum}[1]{\Delta^2_{{#1}}}
\newcommand*{\vev}[1]{\langle{{#1}}\rangle}
\newcommand*{\slprod}[2]{({#1}\mid{#2})}
\newcommand*{\reals}[1]{\mathbb{R}^{{#1}}}

\newcommand*{\modha}{L^{(1)}}
\newcommand*{\modhb}{L^{(2)}}
\newcommand*{\modhab}{L^{(1,2)}}
\newcommand*{\deltat}{\tilde{\delta}}
\newcommand*{\fivegrav}{\kappa_5}
\newcommand*{\fourgrav}{\kappa_4}

\newcommand*{\scale}{\mu}
\newcommand*{\perturbation}{\zeta}

\newcommand*{\efunction}{\mathcal{E}}

\newcommand*{\source}{\mathcal{J}}
\newcommand*{\manifold}[1]{\mathcal{{#1}}}
\newcommand*{\tr}{\mathrm{Tr}}
\newcommand*{\deriv}[2]{\frac{d{{#1}}}{d{{#2}}}}
\newcommand*{\derivtwo}[2]{\frac{d^2{{#1}}}{d{{#2}}^2}}
\newcommand*{\pd}[2]{\frac{\partial{{#1}}}{\partial{{#2}}}}
\newcommand*{\ptwo}[2]{\frac{\partial^2{{#1}}}{\partial{{#2}}^2}}

\newcommand*{\measure}[1]{d{#1}}

\newcommand*{\grad}{\nabla}
\newcommand*{\fmeasure}[1]{[\mathcal{D}{{#1}}]}
\newcommand*{\dthree}[1]{d^3{#1}}
\newcommand*{\KG}{\square_{\text{\textsc{kg}}}}
\newcommand*{\Y}{\square_{\text{\textsc{y}}}}
\newcommand*{\laplacian}{\triangle}

\newcommand*{\bw}{\square_{\text{\textsc{bw}}}}
\newcommand*{\bop}[1]{\mathcal{B}_{{#1}}}
\providecommand*{\arsinh}{\mathrm{arc \; sinh}}

\newcommand{\e}[1]{\mathrm{e}^{{#1}}}

\providecommand*{\eqref}[1]{(\ref{#1})}
\renewcommand{\epsilon}{\varepsilon}

\newcommand{\ba}{\begin{eqnarray}}
\newcommand{\ea}{\end{eqnarray}}
\newcommand{\be}{\begin{equation}}
\newcommand{\ee}{\end{equation}}
\newcommand{\nn}{\nonumber \\}
\newcommand{\x}{\mbox{\boldmath $x$}}
\newcommand{\kb}{\mbox{\boldmath $k$}}

\begin{document}
\title{The consistency relation in braneworld inflation}
\date{\today}
\author{David Seery}
\email{djs@roe.ac.uk}
\author{Andy Taylor}
\email{ant@roe.ac.uk}
\affiliation{Institute for Astronomy, Royal Observatory, Blackford
Hill, Edinburgh, EH9 3HJ, UK}

\pacs{98.80.Cq \hfill astro-ph/0309512}

\preprint{Submitted to Phys. Rev. D}
%
%
\begin{abstract}
The braneworld cosmology, in which our universe is imbedded as a
hypersurface in a higher dimensional bulk, has the peculiar property that the
inflationary consistency relation derived in a four-dimensional
cosmology persists.  This consistency condition relates the ratio
of tensor and scalar perturbation amplitudes to the tensor
spectral index produced during an epoch of slow-roll scalar field
inflation. We attempt to clarify this surprising degeneracy.  Our
argument involves calculating the power spectrum of scalar field
fluctuations around geometries perturbed away from the exact de
Sitter case.  This calculation is expected to be valid for
perturbations which would not cause a late-time acceleration of
the universe. We use these results to argue that the emergence of
the same consistency relation in the braneworld can be connected
with a specific property, that five-dimensional observables
smoothly approach their four-dimensional counterparts as one takes
the brane to infinite tension.  We exhibit an explicit example
where this does not occur, and in which a consistency relation
does not persist.
\end{abstract}
\maketitle
%
%
\section{Introduction}
Recent experimental results from the WMAP project
\cite{wmap} have lent strong support
to the view that the observed homogeneity, isotropy and
small-scale structure of the universe arises from an early period
of accelerated expansion driven by a quantum field, the inflaton,
that violates the weak energy condition. Although other
possibilities exist \cite{starobinsky,holo-inflation}, the
prototypical candidate for a matter component capable of
supporting an inflationary epoch \cite{guth} of this type is a
scalar field.

During inflation, all light fields (mass $m < 3H/2$, where $H$ is the Hubble
parameter during inflation) are excited and pick up a
nearly scale invariant fluctuation whose characteristics are
controlled by the inflaton potential.  In the later universe, the
inflaton fluctuation is communicated \cite{mfb} to the curvature
of spatial slices, seeding primordial structure formation, and
observable today in the large scale distribution of the galaxies
\cite{twodf} and fluctuations in the Cosmic Microwave Background
(CMB). Since the graviton is also massless one would expect small
tensor perturbations to have been excited. Because of the
extremely weak gravitational coupling, these perturbations would
essentially not interact with other constituents of the universe
on their journey towards us and could in principle be observable
today, for example via its imprint in the polarisation field of
the CMB, or with gravitational wave observatories coming on-line
over the next decade. These tensor perturbations would almost
certainly still be in their primordial state and could offer great
insight into the early universe.  In this case one would have four
possible observables, the curvature amplitude,
$\powerspectrum{\perturbation}$, and spectral index,
$n_{\perturbation}$, and the tensor amplitude,
$\powerspectrum{T}$, and spectral index, $n_T$.

Since these quantities are all determined in the scalar field
inflationary model by properties of the scalar potential, one
expects to find some relation between them. In the context of
standard cosmology, one finds \cite{starobinsky-consistency}
 \be
  \label{eq:consistency}
  \frac{\powerspectrum{T}}{\powerspectrum{\perturbation}}
  \approx - 8 n_T,
 \ee
 to lowest order in the slow-roll approximation (see, eg., Ref.
\cite{liddle-lyth}). One can also calculate the next-order term in
the slow-roll expansion \cite{lidsey-liddle}, which does not
preserve the functional form of Eq.~\eqref{eq:consistency}.

Over and above the general current evidence in favour of an
inflationary-like epoch, an observation of this relation in the real
universe would provide extremely strong support for a minimal
scalar field model. More complex models, such as those containing
isocurvature modes, weaken this to an inequality, while observing
an excess of primordial gravitational power would be a severe blow to
the inflationary programme. However, we should stress that the
exact non-perturbative relation between observables is not known;
except in special cases, one does not know how to calculate away
from the slow-roll approximation.

Over the last few years there has been considerable interest in
cosmological models supporting large, extra dimensions
\cite{randall-sundrum-A, randall-sundrum-B}, motivated by
developments in M-theory \cite{horava-witten-a,horava-witten-b}.
It is therefore natural to ask
both how inflation is implemented in these scenarios, and what
possible modifications arise in its predictions for late-universe
observables \cite{maartens-wands,lmw}. Here one discovers a
remarkable surprise.  Although predictions for the tensor and
scalar amplitudes and spectral indices are modified because they
are sensitive to the behaviour of gravity in the large extra
dimensions, the lowest-order consistency relation
Eq.~\eqref{eq:consistency} survives \cite{huey-lidsey-A,
huey-lidsey-B}.  This is a non-trivial feature of the model, and
at the time of writing we are not aware of any simple argument
which demonstrates why this should be true. The unexpected
appearance of the persistence of the consistency relation in the
braneworld potentially jeopardizes the hope of observationally
reconstructing the inflaton potential \cite{liddle-taylor}. Hence
an understanding of the origin of this degeneracy, and in
particular deciding if it is universal, is essential to the
inflaton potential programme.

In this paper, we clarify the circumstances under which one
expects degeneracies between brane cosmology and conventional
cosmology to persist.  The consistency relation,
Eq.~(\ref{eq:consistency}), is derived in the brane and
four-dimensional cases equally by calculating
$\powerspectrum{\perturbation}$ and $\powerspectrum{T}$ using
quantum field theory for the idealization of a perfect de Sitter
cosmology, where the Hubble parameter, $H$, is a constant.  One
then derives $n_{\perturbation}$ and $n_T$ by supposing that $H$
is changing extremely slowly, so that the approximation of a
constant $H$ over any few e-folds of inflation should be good.  We
will argue that the persistence of Eq.~\eqref{eq:consistency} can
be regarded as a particular feature of this model: the tensor
spectrum, $\powerspectrum{T}$, calculated in the brane world joins
smoothly with the four-dimensional result as one decouples the
brane from its surrounding environment. We verify this expectation
by constructing an explicit example containing a small
perturbation away from the exact de Sitter solution.  This perturbation
could be considered as a model of a bulk--brane interaction, or simply
as an imbedded cosmology close to, but not exactly coinciding with, the
de Sitter brane.
One can
successfully calculate the relevant amplitudes and spectral
indices in this model, to first order in the perturbation.
However, one finds that it is not possible to smoothly join the five-%
dimensional result to four-dimensional physics as one takes the brane
to infinite tension.  As a result, one does not recover a consistency
relation.

This paper is organized as follows.  In Section~%
\ref{sec:review} we briefly review the calculation of the 4-D
amplitudes and spectral indices and present a new derivation of
the 5-D braneworld within the framework of Quantum Field Theory
(QFT), both as a convenient reference for the later discussion and
to establish our notation. In Section~\ref{sec:consistency} we discuss the
consistency relation in the four-dimensional case and in the
braneworld. In Sections~\ref{sec:pfourd}--\ref{sec:pfived} we calculate
the effect of an arbitrary density perturbation $\delta\rho(t)$ on
the scalar field power spectrum in the four- and five-dimensional
cases. We begin with an exact de Sitter cosmology fixed by $H =
\text{constant}$, and introduce some small perturbation $\delta
H(t)$. We assume it is still valid to treat the scalar field
fluctuation $\delta\phi$ as a free, massless field propagating on
this background. The two-point function of $\delta\phi$ can then
be calculated both in the brane cosmology and the four-dimensional
universe. In Section~\ref{sec:pconsistency} we apply the results
obtained in Sections~\ref{sec:pfourd}--\ref{sec:pfived} to study
consistency relations in four and five dimensions.  Our strategy
is to write a relationship between observable parameters in the
four dimensional case, and ask whether a comparable relationship
holds in the braneworld.  In fact it will turn out that the
presence of a small perturbation prevents such a relationship.
Finally, we state our conclusions (Section~\ref{sec:conclude}). Some
material extraneous to the main text involving normalization of
the graviton zero mode is presented for reference in an Appendix.
We begin by reviewing the QFT calculation of the power spectra in
four dimensions and presenting a new calculation in the 5-D
braneworld.
\section{The four- and five-dimensional lowest order results}
\label{sec:review}
Scalar field inflation is based on free, massless field theory.
If $\phi$ is the inflaton with some potential $V(\phi)$, then one
treats the gross evolution of $\phi$ classically with the addition of
some fluctuating part $\delta\phi$ which is to be treated quantum
mechanically.  It is a good approximation to take $\delta\phi$ to be
a free, massless field.  The inflaton field $\phi$ itself will not
enter into our considerations, so for the remainder of this paper we
simply drop the $\delta$ from the fluctuating field $\delta\phi$.

\subsection{4-d scalar power spectrum}
\label{sec:fourdscalarpower}
Accordingly, let $\phi$ be a free, massless scalar field.  Its
correlation functions are controlled by the functional integral,
\ba
  \label{eq:correlations}
  \lefteqn{\vev{\phi(x_1) \cdots \phi(x_n)} =}\nn
  & &\int \fmeasure{\phi} \; \phi(x_1) \cdots \phi(x_n) \exp\left(
  -\frac{i}{2}
  \int_{\manifold{M}} \! \measure{x} \; \phi\, \square \,\phi
  \right) ,
\ea
where $\manifold{M}$ is the background spacetime with metric
$g_{\nu \mu}$ and invariant volume measure $\measure{x}$, and we have
chosen units in which $\hbar = 1$.  The operator $\square =
\grad^\mu \grad_\mu$, where $\grad_\mu$ is the covariant
derivative compatible with $g_{\nu\mu}$. One can evaluate the
functional integral in Eq.~\eqref{eq:correlations} explicitly, for example
for the two-point function.  In this case
Wick's theorem shows that $\vev{\phi(x_1)\phi(x_2)} =
-i\square^{-1}(x_1,x_2)$.

Now let $\manifold{M}$ be de Sitter space.  We choose
local coordinates in which the metric takes the form
\begin{equation}
  \measure{s}^2 = \frac{1}{H^2 \tau^2}
  \left( - \measure{\tau}^2 + \delta_{ij} \,
  \measure{x}^i \, \measure{x}^j \right) .
\end{equation}
There are three Killing vectors $\partial / \partial x^i$ which
act transitively on slices $\tau = \text{constant}$. The points
$x_1$ and $x_2$ have coordinates $x_1 = (\tau_1, \x_1)$, $x_2 =
(\tau_2, \x_2)$.  Here, $\x_1$, $\x_2 \in \reals{3}$ but one can
choose the range of $\tau$.  To cover the entire manifold, one
takes $\tau \in \reals{}$.  The spatial slices in this case are
$S^3$.  Alternatively \cite{hawking-ellis}, one can work on the
half space $\tau \in \reals{-} \equiv (-\infty,0]$ which
corresponds to the portion of the manifold covered by the
cosmic-time form of the metric with flat spatial slices,
$\measure{s}^2 = -\measure{t}^2 + \e{2Ht} \delta_{ij} \,
\measure{x}^i \, \measure{x}^j$.  This choice is particularly
common and convenient when discussing inflation.  In either case,
the infinite past corresponds to $\tau \rightarrow -\infty$.  We
work on the full space $\tau \in \reals{}$ because it is easier to
get the boundary conditions right, but in any case $\tau$ usually appears in
the combination $k\tau$ with some wavenumber $k$, and we will take
the large scale limit $k \rightarrow 0$. In this case, one finds
that if $\tau_1 > \tau_2$ the propagator satisfies (see, eg.,
\cite{birrell-davies})
\begin{eqnarray}
  \label{eq:fourd-propagator}
  \vev{\phi(x_1)\phi(x_2)} &=&
   \int_{\reals{3}} \! \frac{\dthree{k}}{(2\pi)^3} \frac{H^2
  \tau_1 \tau_2 \pi}{4k} \modha(-k\tau_1) \nn
  & & \times
  \modhb(-k\tau_2)
  \e{-i\vect{k}\cdot(\x_1 - \x_2)} ,
\end{eqnarray}
where
\begin{equation}
  \label{eq:lfunction}
  \modhab(z) = (z)^{n} H_{3/2}^{(n)}(z) \quad \text{for} \quad n = 1, 2.
\end{equation}
The boundary conditions here are chosen to correspond with the
adiabatic (Bunch-Davies) vacuum prescription \cite[p.
132]{birrell-davies}, that Eq.~\eqref{eq:fourd-propagator} is
close to the flat space limit, $\sim \e{-ik |\tau|}$, whenever the
wavevector is small ($k \rightarrow \infty$) compared to the
curvature of spacetime or when approaching the asymptotically
early or late times, $\tau \rightarrow \pm\infty$. Since $\phi$ is
free, there are no singularities requiring renormalization in the
operator product expansion, although the propagator is log
divergent in both the ultra-violet and infra-red.  One can take
the $\x_1 \rightarrow \x_2$ limit to find an effective variance,
$\sigma^2_\phi(x) = \vev{\phi(x)\phi(x)}$, which satisfies
\begin{equation}
  \label{eq:fourd-power}
  \sigma^2_\phi(\tau)
  = (H \tau)^2 \int_0^\infty \! \frac{k dk}{8\pi} \modha(-k\tau) \modhb(-k\tau) ,
\end{equation}
where $\tau_1 = \tau_2 = \tau$.  This is independent of the spatial
coordinates $\x$.

Throughout this paper the power spectrum, $\powerspectrum{X}$, of
some field, $X$, is defined as
\begin{equation}
  \label{eq:our-ps}
  \powerspectrum{X}(k) = \deriv{\sigma^2_X(k)}{\, \ln k} ,
\end{equation}
where $\sigma^2_X$ is the variance in $X$. The power spectrum
$\powerspectrum{\phi}$ on large scales, $k \rightarrow 0$,
approaches the well-known finite limit
 \be
    \powerspectrum{\phi} = (H/2\pi)^2.
 \ee

The intrinsic curvature perturbation \cite{mfb}
induced by this fluctuation in
the scalar field satisfies $\perturbation = H \delta\phi /
\dot{\phi}$, where $\dot{\phi}$ is the classical background evolution.
Our conventions for this quantity coincide with Ref. \cite{wands-malik}.
Therefore,
\begin{equation}
  \label{eq:fourd-scalar}
  \powerspectrum{\perturbation} =
  (H^2/\dot{\phi}^2)\powerspectrum{\phi}.
\end{equation}

One can now define a sequence of observables based on logarithmic
derivatives of $\powerspectrum{X}$ with respect to $\log k$. The
two lowest members of this sequence are the spectral index, $n_X$,
and its running, $r_X$:
\begin{equation}
  \label{eq:index-running}
  n_X = \deriv{\, \log \powerspectrum{X}}{\, \log k} \quad
  \text{and} \quad
  r_X = \derivtwo{ \log \powerspectrum{X}}{ [\log k]^2} .
\end{equation}

\subsection{4-d tensor power spectrum}
Linear gravitational waves consist of small perturbations $E_{ab}$
to the metric:
 \be
        \measure{s}^2 =
            (\eta_{ab} + E_{ab}) \, \measure{x}^a \,\measure{x}^b.
 \ee
As a representation of the Lorentz group, the metric
perturbation $E_{ab}$ is reducible \cite{weinberg}.
To pick out the pure spin-2 contribution one imposes the
condition that $E_{ab}$ be transverse ($\grad^a E_{ab} = 0$) and
traceless ($\tr \, E_{ab} = 0$) with respect to the Lorentz
metric $\eta_{ab}$.

The gravitational action is
 \be
    S = - \frac{1}{2\fourgrav^2} \int \! \measure{x}
\, R,
 \ee
where $R = \tr\, R_{ab}$ is the trace of the Ricci tensor.
To find an action for $E_{ab}$ one expands $R$ to second order in
$E_{ab}$.  The result is
 \be
    S_2 = - \frac{1}{8\fourgrav^2} \int \! \measure{x} \,
        E^{ab} \square E_{ab},
 \ee
where
$\square = \grad^a \grad_a$ is still the scalar d'Alembertian.
Thus, the Lorentz indices $(a,b)$ of $E_{ab}$ are not noticed and
effectively label an internal $SO(3,1)$, so that $E_{ab}$ behaves
like some number of fields in the trivial (scalar) representation
of the Lorentz group. One includes one such field for each
polarization state of the graviton. In four dimensions, there are
two such polarizations; it is conventional to label one
polarization by $+$ and the other by $\times$. This means that the
action for gravitational perturbations is the same as two copies
of the scalar action, Eq.~\eqref{eq:correlations}, except that the
overall normalization is changed by a factor
$(4\fourgrav^2)^{-1}$.  As a result, the gravitational power
spectrum satisfies
\begin{equation}
  \label{eq:fourd-tensor}
  \powerspectrum{T} = 8 \fourgrav^2 \powerspectrum{\phi} =
  2 \fourgrav^2 \frac{H^2}{\pi^2} .
\end{equation}

\subsection{5-d braneworld}
In the braneword, one works on an anti-de Sitter or
Schwarzschild--anti de Sitter (SAdS) \cite{randall-sundrum-A,
bowcock, global-structure, gubser, verlinde} manifold $\manifold{M}$ with an
imbedded hypersurface $\Sigma$.  Throughout this paper, we
consider only the pure anti-de Sitter case. The hypersurface
$\Sigma$ supports the various matter and gauge fields which
comprise our cosmology.  The metric is taken to be
\cite{binetruy-deffayet-A, binetruy-deffayet-B}
\begin{equation}
  \label{eq:metric}
  \measure{s}^2 = - n^2(t,y) \; \measure{t}^2 + a^2(t,y)
  \delta_{ij} \; \measure{x^i} \, \measure{x^j} + \measure{y}^2 .
\end{equation}
One can replace $\delta_{ij}$ by the metric $\gamma_{ij}$ of any
maximally symmetric three-geometry. The brane is considered to be
imbedded at $y=0$ with a $\zed{2}$ symmetry, and the metric
functions $a(t,y)$ and $n(t,y)$ depend on the
four-dimensional brane geometry.  For this reason, Eq.~\eqref{eq:metric}
is not a product manifold; this is sometimes expressed by saying that
$\manifold{M}$ is a warped compactification.

The $\zed{2}$ symmetry acts by $y \mapsto -y$ and
is motivated from heterotic M-theory
\cite{horava-witten-a, horava-witten-b}.  We loosely refer to this
construction as an orbifold; one can choose either to work on the
full orbifold, or the line interval corresponding to just one half.
In general we will work on the $y > 0$ branch rather than on the full
space; this makes no difference to computations, except that factors of $2$
must occasionally be added by hand.

Except in the case that the brane is empty \cite{randall-sundrum-A},
these coordinates cover only a patch of AdS \cite{global-structure}, so
the coordinate $y$ does not take on unboundedly large values but
instead only assumes values in some interval $y \in [-y_h,y_h]$.
In the $y > 0$ picture, we restrict $y$ to $y \in [0,y_h]$.  The location of
the coordinate horizon at $y = y_h$ depends on the brane tension and
matter theory \cite{lmw}.

The effective Einstein equations on the brane were found in
\cite{shiromizu}.  They are, with $\fourgrav^2$ and
$\fivegrav^2$ the four- and
five-dimensional gravitational couplings respectively,
\begin{equation}
  \label{eq:effective-einstein}
  G_{ab} = \fourgrav^2 T_{ab} + \fivegrav^2 \pi_{ab} + E_{ab} ,
\end{equation}
where $G_{ab}$ is the effective four-dimensional Einstein tensor,
$T_{ab}$ is the energy--momentum tensor of whatever matter
and gauge degrees of the freedom reside on the brane,
$\pi_{ab}$ is quadratic in $T$ and $E$ is the limit as one approaches
the brane of the `electric' part of the Weyl tensor in the bulk.
The hierarchy of Planck scales is controlled by a parameter $\mu$,
\begin{equation}
  \scale = \frac{\fourgrav^2}{\fivegrav^2} = \frac{M_5^3}{M_4^2} .
\end{equation}

The Friedmann equation receives quadratic corrections from the term
$\pi_{ab}$, involving the inverse brane tension $\lambda$
\cite{binetruy-deffayet-B},
\begin{equation}
  \label{eq:brane-friedmann}
  H^2 = \frac{\fourgrav^2}{3} \rho \left(1 + \frac{\rho}{2\lambda} \right) .
\end{equation}
If there is no four-dimensional cosmological constant, then the hierarchy
parameter $\mu$ is the AdS curvature scale and is related to $\lambda$ by
\begin{equation}
  \label{eq:lambda-ell}
  \lambda = \frac{6 \scale}{\fivegrav^2}  .
\end{equation}
For future use, we note that the $\lambda \rightarrow 0$ limit
can be identified with $\scale \rightarrow 0$ at fixed $\fivegrav$,
and similarly as $\lambda \rightarrow \infty$.  The $\lambda
\rightarrow 0$ limit sends $\fourgrav$ to zero and so switches off
four-dimensional gravity on the brane.

The
explicit form of $a(t,y)$ and $n(t,y)$ is, for $a$,
\cite{binetruy-deffayet-B}
\begin{equation}
  \label{eq:a-soln}
  \left( \frac{a}{a_b} \right)^2 = \frac{H^2}{2\scale^2} \left[ \cosh
  2\scale (y_h - y) - 1 \right] ,
\end{equation}
and the conventional choice of $n = \dot{a}/\dot{a}_b$,
\begin{equation}
  \label{eq:n-soln}
  \frac{a_b}{a} n =  1 - \left(
  \frac{a_b}{a} \right)^2 \frac{\dot{H}}{2\scale^2}
  \left( \frac{H^2}{\scale^2 + H^2} \cosh 2\scale(y_{\infty} - y) - 1
  \right)
\end{equation}
and $a_b$ refers to the scale factor on the brane, $a(t,y=0)$.
The quantities $y_h(t)$ and $y_{\infty}(t)$ are defined by,
which are functions of $t$,
 \ba
  \tanh 2\scale y_h &=& \frac{\left(1 + H^2/\scale^2\right)^{1/2}}
  {1+H^2/2\scale^2}, \nn
  \tanh 2\scale y_{\infty} &=& \frac{1}{\left(1 + H^2/\scale^2\right)^{1/2}}
.
 \ea
Clearly, from Eq.~\eqref{eq:a-soln}, $y=y_h$ is always a zero of $a$
and defines a Cauchy horizon or coordinate singularity, where the
Gaussian normal coordiates used to write Eq.~\eqref{eq:metric} break
down.  There is an analytic extension beyond this horizon
\cite{global-structure}.  The location $y=y_h$ is a global minimum
for $a$.
Although $a$ always goes to zero at the Cauchy horizon, in general
$n$ does not; indeed, it is typically discontinuous there.  However
the values of $a$ and $n$ for $y>y_h$ are not meaningful, so this
discontinuity is not seen by observers in the spacetime.  There is
no simple geometric interpretation for $y_\infty$.

\subsection{5-d scalar power spectrum}
Let $\phi$ be a free massless scalar field propagating over
$\Sigma$. Then the propagator for $\phi$ is still defined by
Eq.~\eqref{eq:correlations} (with integration over spacetime $\manifold{M}$
replaced by integration over $\Sigma$) and is exactly the same as
the four-dimensional case, Eq.~\eqref{eq:fourd-scalar}
\cite{wands-malik}.

\subsection{5-d tensor power spectrum }
The situation for gravitational perturbations is more complicated,
and was first analysed by \cite{lmw} (see also Refs. \cite{gorbunov,
giudice-kolb,hawking-hertog})
in the Schr\"{o}dinger picture.
Here we repeat the calculation in a QFT. We will use this approach
to generalize the calculation to an arbitrarily perturbed de
Sitter brane in the Section \ref{sec:pfived}.

Let $E_{ij}$ be a small perturbation of the metric
Eq.~\eqref{eq:metric}:
\begin{equation}
  \measure{s}^2 = - n^2(t,y) \; \measure{t}^2 + a^2(t,y)(\delta_{ij} +
  E_{ij}) \; \measure{x^i} \, \measure{x^j} + \measure{y}^2
\end{equation}
where $E_{ij}$ is transverse and traceless with respect to the
three-dimensional spatial metric $\delta_{ij}$.  Just as in four
dimensions, for the purposes of the resulting field theory, the
indices $(i,j)$ act as internal $SO(3)$ indices and $E$ behaves
like two copies of a field in the trivial (scalar) representation
of $SO(3)$.  Just as in any Kaluza-Klein type decomposition,
in order to make up the full $SO(3,1)$ graviton, one
should include contributions from a graviscalar $\phi$ and
graviphoton $A_i$ which are the other components of a
decomposition of perturbations to the metric under the isometry
group of Eq.~\eqref{eq:metric}.  We ignore $\phi$ and $A_i$; they can
be set to zero by a gauge transformation and do not
contribute to the vacuum fluctuation during inflation
\cite{frolov-kofman}.

The two-point function for $E_{ij}$ satisfies
 \ba
 \vev{E^{ij}(x_1) E_{ij}(x_2)} &=&
    \int \fmeasure{E_{mn}} \; E^{ij}(x_1)
  E_{ij}(x_2) \nn
  & & \hspace{-2 cm}\times   \exp\left[ - \frac{i}{8\fivegrav^2}
  \int_{\manifold{M}} \measure{x} \;
  E^{mn} \left(\frac{\KG}{n^2} + \Y\right) E_{mn} \right]
   ,\nn
 \ea
where we have decomposed the 5-D braneworld d'Alembertian,
$\bw$, into two terms $\KG$ and $\Y$, defined by
 \ba
  \KG &=& - \ptwo{}{t} - \left( 3
  \frac{\dot{a}}{a} - \frac{\dot{n}}{n} \right) \pd{}{t} +
  \frac{n^2}{a^2} \laplacian ,\nn
  \Y &=& \ptwo{}{y} +\left(3 \frac{a'}{a} + \frac{n'}{n} \right)
  \pd{}{y} .
 \ea

Because Eq.~\eqref{eq:metric} is not a product metric, $\KG$ and $\Y$
are not the on- and off-brane d'Alembertians.
$\KG$ is similar to the Klein--Gordon operator
on slices $y = \text{constant}$, but carries an overall
$y$-dependence owing to the fact that both $n$ and $a$ are in general
functions of $y$.
Similarly, $\Y$ depends in general on both $t$ and
$y$.  However, in the important special case that the brane is
endowed with a de Sitter geometry $\dot{H} = 0$, then
these operators separate \cite{lmw}.  In this case $\Y$ is an
honest Sturm--Liouville operator and one can write $E_{ij}$ as a
sum over its harmonics.  This also allows us to re-express the
path integral measure as a product of four-dimensional path
integrals, and we make contact with the four-dimensional physics
using this device.

Following \cite{lmw} we define a set of weighted eigenfunctions,
$\efunction_\alpha(y)$, of $\Y$ by
 \be
    \Y \efunction_\alpha(y) = -(\alpha^2/n^2) \efunction_\alpha(y).
 \ee
In
Sturm--Liouville form this says $(n^4 \efunction_\alpha')' +
\alpha^2 n^2 \efunction_\alpha = 0$.  The $\efunction_\alpha$
can be chosen to be orthonormal in the standard inner product
defined by this measure, ie.,
\begin{equation}
  \label{eq:slnormalize}
  \slprod{\efunction_\alpha}{\efunction_\beta} = 2 \int_0^{y_h}
  \measure{\mu}(y) \; \efunction_\alpha^\ast \efunction_\beta =
  \delta_{\alpha\beta} ,
\end{equation}
where $\measure{\mu}(y) = n^2 \, \measure{y}$,
provided the $\efunction_\alpha$
obey suitable boundary conditions at $y = 0$ and $y = y_h$;
typically such boundary conditions are the Dirichlet--Neumann conditions
$m \efunction_\alpha' + n \efunction_\alpha = 0$, where $m$ and $n$
are independent of $\alpha$.
We have added a factor $2$ by hand in the above equation, to take account
of the other branch of the orbifold.  For physical reasons
\cite{lmw, gorbunov} we choose the derivatives of the
$\efunction_\alpha$ to vanish at $y = 0$ and $y = y_h$. The
allowed values of $\alpha$ consist of a discrete zero-mode bound
state at $\alpha=0$ and a continuum of massive modes for $\alpha >
3H/2$. We now write $E_{ij}$ as an eigenfunction decomposition,
 \be
    E_{ij}(\tau,\x,y) =
        \sum_{\alpha,\omega} E_\omega^\alpha(\tau,\x)\,
    \efunction_\alpha(y)\, \epsilon_{ij}^\omega,
 \ee
where the scalar $E_\omega^\alpha(\tau,\x)$ is independent of $y$,
and $\epsilon_{ij}^\omega$ is a constant $SO(3)$ polarization
tensor labeled by an index $\omega$.  To take advantage of this
decomposition we rewrite the path integral measure as
\begin{equation}
  [ {\cal D} E_{ij}] \propto
  \prod_{\alpha, \omega} [ {\cal D} E_\omega^\alpha] ,
\end{equation}
and the two-point function becomes
 \ba
  \lefteqn{\vev{E^{ij}(x_1,y_1) E_{ij}(x_1,y_2)} =} \nn
    & & \int\! \prod_{\alpha,\omega}
  \fmeasure{E_\omega^\alpha} \,\, \left(
  \sum_{\alpha', \alpha'', \omega'}
  E_{\omega'}^{\alpha'} E_{\omega'}^{\alpha''}
  \efunction_{\alpha'}(y_1)
  \efunction_{\alpha''}(y_2) \right) \nn
  & & \label{eq:braneworldgravaction} \times
  \exp\left(- \frac{i}{8\fivegrav^2} \int_\Sigma dx \;
  \sum_\omega
  E_\omega^\alpha ( \KG  - \alpha^2) E_\omega^\alpha \right)
  .
 \ea
We have integrated over the transverse dimension, so
all factors of $n$ have disappeared from the measure and $\KG$;
cf. Eq.~\eqref{eq:slnormalize}.
Thus the field $E_{ij}$ behaves like a collection of
four-dimensional Klein--Gordon fields with masses described by the
allowed values of $\alpha$. At low energies, or during inflation,
only the $\alpha = 0$ zero-mode will be excited, so since
$\efunction_0$ is independent of $y$, one has
 \ba
  \vev{E^{ij}(x_1) E_{ij}(x_2)} &\approx&
  \efunction_0^2 \sum_{\omega'} \int
  [{\cal D} E^0_\omega] \; E^0_{\omega'} E^0_{\omega'} \nn
  & & \times \exp\left(-
  \frac{i}{8\fivegrav^2} \int_\Sigma dx \;
  \sum_\omega E^0_\omega \KG  E^0_\omega\right) \nn
  &=& -8i \efunction_0^2 \fivegrav^2 \KG^{-1}(x_1,x_2) .
 \ea
This is still a free theory, so there is no obstruction to taking
the limit $x_1 \rightarrow x_2$ and the power spectrum follows in
the same way as the four-dimensional case. When the imbedded
geometry is purely de Sitter $n$ is independent of $t$, and $a =
\e{Ht} n$, so $\KG$ is the same operator as $\square$ on 4-D de
Sitter space.  Re-using the result Eq.~\eqref{eq:fourd-power}, one
obtains
\begin{equation}
  \powerspectrum{T} = 8 \efunction_0^2 \fivegrav^2
  \powerspectrum{\phi} = 2 \efunction_0^2 \fivegrav^2
    \frac{H^2}{\pi^2} .
\end{equation}
The polarizations are labeled by $\omega \in \{ + , \times
\}$ as in four dimensions. Because the $\efunction_0$ are just
constants, this is no more than a renormalization of the
four-dimensional power spectrum by some function of $H$ and
$\scale$.  In order to make this property more transparent, it is
conventional to re-write the power spectrum in terms of the
effective four-dimensional gravitational coupling $\fourgrav^2$ by
setting $\efunction_0^2 = \mu F^2$, where $F$ is a constant to be
fixed by the normalisation of $\efunction_0$ (see Appendix). With
these conventions, one has the simple result,
\begin{equation}
  \label{eq:fived-tensor}
  \powerspectrum{T} = 2 \fourgrav^2 F^2 \frac{H^2}{\pi^2}.
\end{equation}

\section{The consistency relation}
\label{sec:consistency}
We now briefly describe how the consistency relation,
Eq.~\eqref{eq:consistency}, arises in four dimensions, and in the
braneworld. Consider four-dimensional inflation driven by some
scalar field.  The matter and tensor power spectra satisfy
Eqs.~(\ref{eq:fourd-scalar}) and (\ref{eq:fourd-tensor}).
Therefore, their ratio is independent of $\powerspectrum{\phi}$:
\begin{equation}
  \label{eq:fourd-ratio}
  \frac{\powerspectrum{T}}{\powerspectrum{\perturbation}} = 8
  \fourgrav^2
  \frac{\dot{\phi}^2}{H^2} .
\end{equation}
This arises because $\phi$ and $E_{ab}$ share the same action, and
the result is that the ratio Eq.~\eqref{eq:fourd-ratio} depends
only on the relation between $H$ and $\phi$, and not on the
details of the quantum field theory calculation leading to
$\powerspectrum{\phi}$.  If one assumes that the scalar field
$\phi$ is the only constituent of the universe, then $H$ evolves
with $\phi$ according to the equation
\begin{equation}
  \label{eq:fourd-hj}
  H' = - \frac{\fourgrav^2}{2} \dot{\phi} .
\end{equation}
A prime denotes a derivative with respect to $\phi$. This is often
called the Hamilton--Jacobi equation \cite{lidsey-liddle}.  One
then eliminates $\dot{\phi}$ from the ratio
Eq.~\eqref{eq:fourd-ratio} to find
\begin{equation}
  \frac{\powerspectrum{T}}{\powerspectrum{\perturbation}} = \frac{32}
  {\fourgrav^2} \left( \frac{H'}{H} \right)^2 .
\end{equation}
We define a tensor spectral index $n_T$ as in
Eq.~\eqref{eq:index-running}. To evaluate this we endow $H$ with
some extremely slow time dependence owing to motion of $\phi$
on very long time scales.  One quantifies this time dependence by
introducing slow roll parameters $\epsilon$, $\eta$
\cite{liddle-lyth} defined by
\begin{equation}
  \epsilon = \frac{2}{\fourgrav^2} \left( \frac{H'}{H} \right) \quad
  \text{and} \quad
  \eta = \frac{2}{\fourgrav^2} \frac{H''}{H}
\end{equation}
and demanding that these are both small, $\epsilon, \eta \ll 1$.
One typically works to first order in $\max \{ \epsilon, \eta \}$.

On horizon scales $k
\approx aH$, the differential $\measure{\, \log k}$ is effectively
$\measure{\, \log a}$ since we are assuming that change in $H$ is
negligible in comparison with change in $a$.
This approximation is expected to be good because $a = \e{Ht}$ is moving
exponentially fast in comparison with $H$, which assumed almost
constant.  We use
this result to evaluate $n_T$.

With these choices, one obtains Eq.~\eqref{eq:consistency}. This
result depends on the relationship between $\dot{\phi}$ and $H$
and the functional form of $\powerspectrum{\phi}$.  In particular,
if $\powerspectrum{\phi}$ is just monomial in $H$,
ie., $\powerspectrum{\phi} \propto H^n$ for some $n$,
and $\dot{\phi}$ is
linear in $H'$, then one will obtain a consistency relation of
this type, with coefficient depending on
the details of the relationships.

\subsection{The consistency relation in the braneworld}

In the braneworld, the ratio
$\powerspectrum{T}/\powerspectrum{\perturbation}$ is still
independent of $\powerspectrum{\phi}$. One can again write a
relation between $\dot{\phi}$ and $H$, which is the analogue of
the four-dimensional Hamilton--Jacobi equation.  This is modified
by the anti-de Sitter radius $\scale$, and becomes
\begin{equation}
  \label{eq:fived-hj}
  \dot{\phi} = - \frac{2\scale}{\fourgrav^2} \frac{H'}{(H^2 +
  \scale^2)^{1/2}} .
 \ee
This means that the ratio of the amplitudes satisfies
\begin{equation}
  \label{eq:spectra-ratio-fived}
  \frac{\powerspectrum{T}}{\powerspectrum{\perturbation}} = \frac{32
  \scale^2}{\fourgrav^2} F^2 \left( \frac{H'}{H} \right)^2
  \frac{1}{H^2 + \scale^2} .
\end{equation}
The tensor spectral index no longer depends merely on the functional
form of $\powerspectrum{\phi}$, which is unchanged from the
four-dimensional case, but instead receives non-trivial
corrections from the
renormalization $F^2$.  It satisfies $\measure{\, \log
\powerspectrum{T}} = 2 \measure{\, \log HF}$.  However, $HF$
satisfies a particular differential equation \cite{huey-lidsey-A}:
\begin{equation}
  \label{eq:odd-diff-eq}
  \measure{\, \log HF} = \frac{\scale F^2}{(H^2 + \scale^2)^{1/2}} \;
  \measure{\, \log H} .
\end{equation}
Combining $n_T$ with Eq.~\eqref{eq:fived-hj} and
Eq.~\eqref{eq:odd-diff-eq} gives back the consistency relation
Eq.~\eqref{eq:consistency}. The relation
Eq.~\eqref{eq:odd-diff-eq} is derived, together with an explicit
expression for $F$, in Appendix \ref{sec:diff-eq}.

Although we have derived this result only for the case of a pure
anti-de Sitter bulk with $\zed{2}$ symmetry, the appearance of the
consistency relation holds rather more generally.  In particular,
Huey \& Lidsey \cite{huey-lidsey-A} have argued that it persists if one
allows
different anti-de Sitter curvatures $\scale_<$, $\scale_>$ on
the $y<0$ and $y>0$ branches.

\subsection{The large $\lambda$ limit}

This coincidence appears remarkable.  We have non-trivially
modified gravity in the brane world scenario, but when one asks
about relationships between observable parameters, the
modification becomes invisible.  For a deeper understanding of
brane cosmology in general and to clarify the observational
position, one would like to understand how this circumstance
arises. In particular, one would like to know whether its
appearance in this model is in some sense accidental, or is
enforced by deeper principles. Our claim is that its appearance is
only accidental.

The appearance of the consistency relation Eq.~\eqref{eq:consistency} in
the brane universe theory depends on the differential equation
Eq.~\eqref{eq:odd-diff-eq}, for which it is difficult to find any
clear geometrical or physical interpretation.  Its effect is to render
the relationship between $\powerspectrum{\perturbation}$,
$\powerspectrum{T}$ and $n_T$ independent of the brane
tension $\lambda$ which appears in the Friedmann equation.
One approach to the problem is
to consider the limit of very large $\lambda$, where the brane is
close to decoupling from the bulk and one is nearly in the regime of
four-dimensional cosmology.  Physics on the brane is supposed to be
insensitive to the value of $\lambda$, which we formalize by demanding
that the energy--momentum tensor of matter carried by the brane
is independent of $\lambda$.  If this is true, then the density $\rho$
does not depend on $\lambda$ either and the only appearance of the
tension is in the Friedmann equation, as explicitly written in
Eq.~\eqref{eq:brane-friedmann}.

In fact this is part of a more general correspondence principle,
that one should recover four dimensional physics as the brane
becomes infinitely stiff and decouples from the bulk.  This is true
for purely four-dimensional quantities, such as observables in any
quantum field theory fixed to the brane, but it need not be true
for gravitational quantities.  Consider, for example, the
effective Einstein equations Eq.~\eqref{eq:effective-einstein}.
As $\lambda \rightarrow \infty$, the quadratic term $\pi$ which leads
to the quadratic $\rho$ dependence in Eq.~\eqref{eq:brane-friedmann}
decouples,
but $E_{ab}$ need not.  This means that unless $E_{ab} = 0$, then as
one decouples the brane one need
not see gravitational quantities such as
$\powerspectrum{T}$ and $n_T$ computed using
Eq.~\eqref{eq:effective-einstein} approach their counterparts
computed in four dimensional Einstein gravity.

For this reason, a useful way to look for models which might break
consistency relation-type degeneracies between brane and four-dimensional
cosmologies is to search for examples in which observable quantities
do not smoothly approach their four-dimensional Einstein counterparts
as one decouples the brane.  In the next section, we will examine
such a model.

To decide how the limit $\lambda \rightarrow \infty$ ought to be taken,
one needs some sort of prescription for handling the matter on $\Sigma$.
For example,
the Einstein--Klein--Gordon system for a four-dimensional Robertson--Walker
universe or an imbedded Robertson--Walker universe in the five-dimensional
case can be written as a pair of equations for the scalar field,
\begin{equation}
  \rho = \dot{\phi}^2/2 + V, \quad \text{and} \quad \rho' = - 3H \dot{\phi},
\end{equation}
supplemented with an equation for the Hubble parameter,
Eq.~\eqref{eq:brane-friedmann} in the 5d brane case
or $H^2 = \fourgrav^2 \rho / 3$ in the 4d case.
Once one specifies any one
of these quantities in terms of $\phi$, the other two are determined
by direct differentiation or quadrature, as appropriate.

de Sitter space corresponds to a fixed value of $H$.  In four dimensions,
this translates straightforwardly to a fixed $\rho$ and a fixed $V$.
In five dimensions, the situation is not quite so simple; in particular,
as the tension $\lambda$ varies, one must vary $\rho$ and $V$ to keep
$H$ fixed.  Therefore, if one wishes to deal with some given
four-dimensional cosmology and then imbed this as a brane universe, one
must specify which characteristic, selected from $H$, $\rho$ or $V$, is
to be used to determine the brane geometry.  The other two will vary
with $\lambda$.  In the present case, we have a single fixed $H$.
If one now applies a perturbation $\delta H(t)$, then we fix our notion
of what is the same cosmology imbedded on a brane by asking that one
obtains the same perturbation $\delta H(t)$.  This must be sourced
by a perturbation $\delta\rho(t,\lambda)$ which is also a function of
$\lambda$.
\section{Fluctuations in a perturbed 4-D de Sitter space}

\label{sec:pfourd}

In this section, we aim to calculate the power spectrum of a
scalar field propagating over a background de Sitter cosmology
with some first order perturbation. Let $M$ be four-dimensional de
Sitter space with Hubble parameter $H_0$.  Consider a matter
density perturbation $\delta\rho$ of square zero which induces a
change $\delta H$ in the Hubble parameter. We wish to the
calculate the power spectrum of a free, massless scalar field
propagating on this geometry.

The functional integral Eq.~\eqref{eq:correlations} defining
correlation functions in the scalar field theory is unchanged, so
the two point function is still given by $-i \square^{-1}$,
although $\square$ now includes contributions of $O(\delta H)$
from the perturbation. For convenience we adopt the convention of
writing the inverse d'Alembertian as a Green's function; $G =
\square^{-1}$. Then $G$ satisfies
\begin{equation}
  \label{eq:perturb-scalar-field}
  \square \,G(x_1,x_2) = \delta(x_1-x_2)
\end{equation}
where $\delta(x_1-x_2)$ is the covariant Dirac delta function,
which can be written in terms of the coordinates $\x_1^i$,
$\x_2^i$ of $\x_1$ and $\x_2$ as $(-\det g)^{-1/2} \prod_i
\delta(\x_1^i - \x_2^i)$. Although the Green's function is
symmetric between $\x_1$ and $\x_2$ it is helpful in calculations
to adopt the convention that $G$ is a function of one set of
coordinates, for which we choose the $\x^i$.  The coordinates
$\x^i$ are then considered to be constants. $G$ can be solved as a
perturbation expansion in $\delta H$. We write $G = G_0 + \delta
G$, and because there still exist the three spacelike Killing
vectors $\partial /
\partial x^i$ it is useful to diagonalize $G$ by writing it
as a Fourier transform, $G(\x,\x_2) = \int \dthree{k} \; (2\pi)^{-3}
\; G(\kb; t, t_2) \e{-i\kb\cdot(\x - \x_2)}$. The $O(1)$
equation for $G_0(\kb)$ is
\begin{equation}
  \left( \ptwo{}{t} + 3H_0 \pd{}{t} + \frac{k^2}{a^2} \right)
  G_0 = \frac{\delta(t - t_2)}{a^3} .
\end{equation}
Here the independent variable $t$ on the left hand side is taken
to be the $t$ coordinate of $x_1$, as discussed above.  We still
write the $t$ coordinate of $x_2$ as $t_2$.  By making the
transition to conformal time $\tau$ defined by $\measure{t} = a \,
\measure{\tau}$ and rescaling $G_0 \rightarrow  a G_0 $, this
becomes
\begin{equation}
  \label{eq:u-unperturb}
  \bop{3/2} G_0 = \frac{\delta(\tau - \tau_2)}{a^3} .
\end{equation}
The operator $\bop{\mu}$ on the left-hand side will occur again, so it is
convenient to have a notation for it.  It is defined by
\begin{equation}
  \bop{\mu} = \ptwo{}{\tau} + \left( k^2 - \frac{\mu^2 - 1/4}{\tau^2}
  \right) .
\end{equation}
The function $G_0$ is almost everywhere zero and so should lie in the
space $\ker \bop{\mu}$ of functions annihilated by $\bop{\mu}$.
This is a two-parameter space of functions spanned by linear combinations
of the form
\begin{equation}
  \Upsilon (-k\tau)^{1/2} H_\mu^{(1)}(-k\tau)
  + \Xi (-k\tau)^{1/2} H_\mu^{(2)}(-k\tau)
\end{equation}
for some constants $\Upsilon$ and $\Xi$.
The precise linear combination one chooses for $G_0$ depends on the boundary
conditions in the far past ($\tau \rightarrow -\infty$) and the far
future ($\tau \rightarrow +\infty$).  A common choice is the Bunch--Davies
vacuum \cite{birrell-davies}, where one chooses $G_0$ to behave like
the Hankel function
$H^{(2)}$ near $\tau \rightarrow +\infty$, and like $H^{(1)}$ near
$\tau \rightarrow -\infty$.  Demanding that $G_0$ be continuous at $\tau
= \tau_q$, but with a step in derivative to satisfy
Eq.~\eqref{eq:u-unperturb} gives back the four dimensional propagator
quoted as Eq.~\eqref{eq:fourd-propagator}.

The $O(\delta H)$ equation is
 \ba
  \lefteqn{\left( \ptwo{}{t} + 3 H_0 \pd{}{t} + \frac{k^2}{a^2} \right)
  \delta G =} \nn
  & &  \left( \frac{2k^2}{a^2} \frac{\delta a}{a}
  - 3 \delta H \pd{}{t} \right)  a G_0
   -
  3 \frac{\delta a}{a} \frac{\delta(t - t_2)}{a^3} .
 \ea
One now follows the same procedure as above, writing $\delta G
\rightarrow a \delta G$ and changing to conformal time. The result
is
\begin{eqnarray}
  \bop{3/2} \delta G & = & 2 \left[k^2
  G_0 \frac{\delta a}{a} - 3 \delta H \left( a \pd{G_0}{\tau} - a^2 H_0 G_0
  \right)\right] \nn
  & &  - 3 \frac{\delta a}{a} \frac{\delta(\tau - \tau_2)}{a} \\
  \label{eq:perturb-fourd}
  & = & \source(\tau,\tau_2) - 3 \frac{\delta a}{a} \frac{\delta(\tau -
  \tau_2)}{a} .
\end{eqnarray}
We have written the non-distributional part of the source term, in
square brackets on the right-hand side, as $\source(\tau,\tau_2)$.
This is a useful abbreviation, but in any case it is
convenient to work with a quite general source term because we
will be able to reuse the result in \S\ref{sec:pfived} below.

$\bop{\mu}$ is a well-defined Sturm--Liouville operator for any
$\mu$ except at $\tau = 0$. We define a set of eigenfunctions $\{
\phi_m \}$ of $\bop{3/2}$ by
 \be
    \bop{3/2} \phi_m = - m^2 \phi_m,
 \ee
which is supplemented by appropriate boundary conditions.  We take the $\{
\phi_m \}$ to be defined on (at least) $\tau \in \reals{-}$, so
the boundary condition at $\tau = -\infty$ is expected to be
immaterial, provided the $\{ \phi_m \}$ decay sufficient fast
there.  At $\tau = 0$, we demand that the $\{ \phi_m \}$ be
regular.  Then a standard argument shows that $\bop{3/2}$ is
self-adjoint on the $\{ \phi_m \}$ and consequently that we may
choose the $\{ \phi_m \}$ to be orthonormal for different $m$.
Explicitly, the appropriately normalized $\{ \phi_m \}$ satisfy
\begin{equation}
  \label{eq:define-phim}
  \phi_m(k,\tau)  = (-\sqrt{k^2+m^2} \tau)^{1/2} J_{3/2}(-\sqrt{k^2+m^2} \tau),
\end{equation}
the Bessel function $J_{3/2}$ being chosen to keep the $\phi_m$ regular
at $\tau = 0$.

Our strategy is to solve Eq.~\eqref{eq:perturb-fourd} by taking a
transform in the $\{ \phi_m \}$.  This can be thought of as the
continuum limit \cite{morse-feshbach}
of an eigenseries expansion in the $\{ \phi_m \}$, cut
off at some limiting value $\tau = - \tau_{\text{limit}}$, as
$\tau_{\text{limit}} \rightarrow \infty$.
One writes the term $\source$ as a $\bop{3/2}$--transform,
\begin{equation}
  \source(k,\tau,\tau_2) = \int_{-\infty}^{\infty} \!\measure{m}
  \,
  \phi_m(k,\tau) \int_{-\infty}^0 \!\measure{\eta} \, \phi_m(k,\eta)
  \source(k,\eta, \tau_2) .
\end{equation}
By inspection of Eq.~\eqref{eq:define-phim} it can be seen that in
fact this is no more than the Fourier--Bessel representation of
$\source(\eta, \tau_q)$. We assume a solution is possible of the
form $\delta G(\tau) = \int_{-\infty} ^{\infty} \measure{m} \,
\phi_m(\tau) \delta G(m) + $ elements of $\ker \bop{3/2}$, where
we indicate whether $\delta G(\tau)$ or some component of its
transform $\delta G(m)$ is under discussion by writing in the
argument explicitly. Substituting into
Eq.~\eqref{eq:perturb-fourd} allows one to solve exactly for
$\delta G(\tau)$:
 \ba
  \delta G(k,\tau) &=& \int_{-\infty}^{\infty}\!
  \frac{\measure{m}}{-m^2}\,
  \phi_m(k,\tau) \int_{-\infty}^0 \!\measure{\eta}\, \phi_m(k,\eta)
  \source(k,\eta, \tau_2) \nn
  & & - 3 \left.\frac{\delta a}{a}\right|_{\tau_2}
  G_0(\tau, \tau_2) .
 \ea
The term $G_0 \in \ker \bop{3/2}$ is chosen to represent the $\delta$-%
function in Eq.~\eqref{eq:perturb-fourd} because it trivially
gives back the same behaviour near $|\tau| \rightarrow \infty$ as
$G_0$. We are also writing $\source(\tau,\tau_2)$ only over the
range $\tau \in \reals{-}$, because it is convenient to define the
source function in the $t$-frame where $\source(\tau,\tau_2)$ is
then undefined for $\tau > 0$.

This procedure makes sense provided the Fourier--Bessel transform
of $\source$ exists; to check convergence of the integral, it is
necessary to assess the behaviour of $\source(\eta,\tau_2)
\phi_m(\eta)$ both as $\eta \rightarrow -\infty$ and $\eta
\rightarrow 0$.

In the $\eta \rightarrow -\infty$ case, the $\phi_m(\eta)$ tend to
oscillating functions of $\eta$.  In this case the integral converges
provided $\delta\rho \rightarrow 0$ as $\eta \rightarrow -\infty$.
The $\eta \rightarrow 0$ case is more complicated.  In this limit,
$\phi_m(\eta) \sim \eta^2$, whereas $G_0(\eta) \sim \eta^{-1}$.
The first term in $\source$ diverges like $G_0 \delta a / a$, and
since $\delta a / a \sim \delta \rho / \rho$ this term behaves near
$\eta \rightarrow 0$ like $a^{-1} \delta \rho$.  To prevent a divergence,
$\delta \rho$ must not diverge faster than $a$.  Using the
thermodynamic redshifting law $\rho \sim a^{-3(1+w)}$, where $p = w\rho$
is the equation of state, this translates into an asymptotic
equation of state stiffer than $w = -4/3$.  This is satisfied for
all forms of matter obeying causal propagation \cite{hawking-ellis}.

The remaining two terms behave like $a \delta \rho$, for which
$\delta \rho$ must go to zero faster than $a^{-2}$ where the integral
would be logarithmically divergent.  This gives
$w > -1/3$, which precludes any form of matter leading to accelerated
expansion or quintessence-like behaviour in the late universe, but allows
any form of normal matter with $w \geq 0$.  In particular, if one
imagines $\delta \rho$ dying away in the infinite future (in the
$t$-frame) as well as
the infinite past, then the integral will be well behaved.  There are no
strong
restrictions on the behaviour of $\delta \rho$ in the intermediate
region between the asymptotic past and future.

One can now assemble $G_0$ and $\delta G$ to construct the full
two-point function, restoring the necessary factors of
$\e{i\vect{k}\cdot \vect{p}}$ and $a$, and integrations over $\vect{k}$. We
let $\x_1$ approach $\x_2$, which gives
 \be
 \label{eq:fourd-fulltwopoint}
  -i G(\x_1,\x_1) =   H_0 \tau \int \! \frac{d^3\!k}{(2\pi)^3}\,
    W(k,\tau,\tau_2)
 \ee
 where
 \ba
    \lefteqn{W(k,\tau,\tau_2) =}& &  \nn
    & &
            \frac{\pi}{4k} H_0 \tau_2 \left(1 - 3 \frac{\delta
  a}{a} \right)
  \modha(-k\tau_2) \modhb(-k\tau) \nn
  & &
   +  i
  \int_{-\infty}^{\infty}\! \frac{\measure{m}}{m^2} \,\phi_m(k,\tau)
  \int_{-\infty}^0 \! \measure{\eta} \, \phi_m(k,\eta)
  \source(k,\eta,\tau_2) ,
  \ea
if $\tau > \tau_2$ and the same expression with $H_{3/2}^{(1)}$ and
$H_{3/2}^{(2)}$ interchanged if $\tau < \tau_2$.
$\modhab(z)$ is defined by Eq.~\eqref{eq:lfunction} in
Section~\ref{sec:fourdscalarpower}, as before.
To find the
power spectrum one lets $\tau$ approach $\tau_2$ and takes a
logarithmic derivative with respect to $k$.  The result is
\be
  \label{eq:fourd-perturb-ps}
  \powerspectrum{\phi} = \frac{4\pi k^3}{(2\pi)^3} H_0 \tau W(k,\tau,\tau) . \ee To take the $k \rightarrow 0$ limit, one
needs to know the behaviour of $\source(\eta,\tau)$ at small $k$.
The term proportional to $\delta H$ vanishes, because $\partial
G /
\partial t$ vanishes on large scales, leaving only the first term
which comes from perturbing the Laplacian.  As a result,
\begin{equation}
  \label{eq:source-asymptotics}
  \source(\eta,\tau) \overset{k \rightarrow 0}{\longrightarrow}
  H_0\tau \frac{\pi i k}{2} \frac{\delta a}{a}(\eta)
  \frac{2^3 \Gamma^2(2/3)}{\pi^2 (-k\eta) (-k\tau)} .
\end{equation}
Substituting this into the limiting form of
Eq.~\eqref{eq:fourd-perturb-ps} gives
 \ba
  \label{eq:perturb-fourd-spectrum}
  \powerspectrum{\phi} &=& \left( \frac{H_0}{2\pi} \right)^2 \times \nn
  &  & \!\!\!\left(1 - 3
  \frac{\delta a}{a} - \frac{2}{\tau}
  \int_{-\infty}^{\infty} \!\frac{\measure{m}}{m^2}\,
  \hat{\phi}_m(\tau)\int_{-\infty}^0 \!\! \frac{d\eta}{\eta}
  \,
  \hat{\phi}_m(\eta) \frac{\delta a}{a}(\eta)
  \right) ,\nn
 \ea
where we define a new set of functions
$\{ \hat{\phi}_m \}$ to be the $k \rightarrow 0$ limit of the
eigenfunctions $\{ \phi_m \}$.  To achieve this result, we have evaluated
the answer on the horizon scale $k = aH = -1/\tau$.
\section{Gravitational fluctuations in a perturbed braneworld}
\label{sec:pfived}
We now aim to repeat this calculation in the
braneworld.  As before, we consider a de Sitter brane with Hubble
parameter $H_0$ immersed in anti-de Sitter space and allow small
fluctuations $\delta\rho$ in the matter density.  However as
discussed above these fluctuations are taken to vary with
$\lambda$ in such a way as to keep $\delta H$ the same.  We define
this to be our notion of the `same' perturbation in the brane
world and in four dimensions.

Firstly, consider some scalar field $\phi$ propagating over the
brane $\Sigma$.  The operator $\square$ appearing in the scalar
field action is the same as would arise in the four-dimensional
case, so the theory is the same as Eq.~\eqref{eq:perturb-scalar-field}
and the resulting power spectrum satisfies
Eq.~\eqref{eq:perturb-fourd-spectrum}, provided that $a$ is taken to
satisfy the expansion law for the on-brane cosmological scale
factor.  One would relabel this quantity $a_b$ in the brane case.

The case of gravitational waves is not the same. In a general
geometry, the graviton wave operator $\bw$ couples the
$t$ and $y$ dependence of the graviton $\vect{k}$-modes, so that
an explicit solution is extremely difficult. One can always work
on the brane universe in black hole coordinates
\cite{global-structure,bowcock}, where the metric is explicitly
stationary, and one recovers ordinary differential equations.
Unfortunately, the boundary conditions are non-trivial to apply.
In this section we make progress by a different route.

We begin by rewriting the general formula Eq.~\eqref{eq:n-soln}
for $n(t,y)$ in terms of $H'$, where
 \be
    \dot{H} = \dot{\phi} H' =
          - \frac{2\scale}{\fourgrav^2}
          \frac{H'^2}{\sqrt{H^2+\scale^2}}.
 \ee
Since $\dot{H} \propto H'^2$, if we perturb around the de Sitter
solution, the term $H' = \delta H$ squares to zero.
Hence, for a perturbed de Sitter brane, we still retain
$n=a/a_b$. We emphasize that this is only true for perturbations around de
Sitter space supported by a scalar field where the background $H$
satisfies $H' = 0$. When
calculating spectral indices we will again endow $H$ with very
weak time dependence, but for the purposes of the calculation
presented in this section the background $H$ is to be regarded as
fixed, in analogy with the four dimensional calculation of
$\powerspectrum{\phi}$.

Let us write the metric functions $a$ and $n$ as in
Eq.~\eqref{eq:a-soln}--\eqref{eq:n-soln},
\begin{equation}
  \label{eq:nunperturbed}
  n^2(y) = \frac{H^2}{2\scale^2} \left[ \cosh 2 \scale(y_h - y) - 1
  \right]
\end{equation}
and $a(t,y) = a_b(t) n(y)$
where $a_b(t)$ is the scale factor on the brane.
Under a variation $H \mapsto H_0 + \delta H$, the function $n(t,y)$
becomes
\begin{equation}
  n(t,y) \mapsto \left(1 + \frac{\delta H}{H_0} \right)
  \frac{H_0}{\sqrt{2} \scale}
  \left[ \cosh 2 \scale(y_h + \delta y_h - y) - 1 \right]^{1/2} ,
\end{equation}
since the horizon location $y_h$ in principle depends on time.
However, the variation in $y_h$ can be safely ignored, because the small
term $\delta y_h$ vanishes inside $\cosh$, that is,
$\cosh (X_0 + \epsilon) = \cosh X_0 + O(\epsilon^2)$
for any $X_0$ and small quantity $\epsilon$.  Therefore, one may still
take the effective horizon to sit at $y=y_h$ and write
\begin{equation}
  n(y) \mapsto (1+\deltat) n_0(y), \quad \deltat = \frac{\delta H}{H_0}
\end{equation}
where $n_0(y)$ is the unperturbed function Eq.~\eqref{eq:nunperturbed}.
The metric is $\measure{s}_5^2 = n_0^2 \, \measure{s}_4^2 + \measure{y}^2$,
where
\begin{equation}
  \measure{s}_4^2 = (1 + \deltat)^2\left[ -\measure{t}^2 +
  (a_b + \delta a_b)^2 \delta_{ij} \, \measure{x}^i \, \measure{x}^j
  \right] .
\end{equation}

This guarantees that $\KG$ and $\Y$ separate, and
under these circumstances $\Y$ depends only on $y$.
The operators $\KG$ and $\Y$ take the explicit form
 \ba
  \KG &=& -\ptwo{}{t} - \left(3 H + 2 \pd{\deltat}{t} \right)
  \pd{}{t} + \frac{\laplacian}{a_b^2}, \nn
  \Y &=& \ptwo{}{y} + 4 \frac{n_0'}{n_0} \pd{}{y} .
 \ea
The operator $\Y$ is again an honest Sturm--Liouville
operator. We consider weighted eigenfunctions of the form
 \be
    \Y \efunction_\alpha = - (\alpha^2 / n_0^2) \efunction_\alpha.
 \ee
The $\efunction_\alpha$ are chosen by fixing the derivatives
$\efunction_\alpha'$ to vanish at $y = 0$ and $y = y_h$; with this
choice, they can be made orthogonal in the inner product
$\slprod{\efunction_\alpha}{\efunction_\beta} = \int
\measure{\mu}(y) \; \efunction_\alpha^\ast \efunction_\beta$ as
before, where $\measure{\mu}(y) = n_0^2 \, \measure{y}$. In
particular, the normalization $F$ of the $\efunction_0$
eigenfunctions depends only on $H_0$ and not $\delta H$, because
$\Y$ does not see the perturbations: it is the same as the unpertuebed
operator.
The gravitational action, after decomposing into $SO(3)$ polarizations,
is equivalent to a certain number of copies of the scalar field
action,
one for each polarization state of the graviton.
The action is (cf. Eq.~\eqref{eq:braneworldgravaction})
\begin{equation}
  \label{eq:perturb-grav-action}
  \frac{1}{2}
  \int_{\Sigma} \! \measure{x}  \, \sum_\alpha E^\alpha \left(
  \frac{\KG}{(1+\deltat)^2} - \alpha^2 \right) E^\alpha
\end{equation}
where the integration is in the metric $\measure{s}_4^2$ and all
functions $n_0(y)$ have disappeared.

The important feature here is that Eq.~\eqref{eq:perturb-grav-action} is
not the same as the action for a perturbed four-dimensional scalar
field.  Some structure left over from the higher-dimensional graviton
operator $(\KG/n^2) + \Y$ is still visible, which modifies the result.

Consider the massless $\alpha = 0$ mode.  Just as in the
unperturbed case, this is the important contribution for
perturbations generated during inflation.  The higher
Kaluza--Klein modes with $\alpha > 3H/2$ are heavy in the sense
that their fluctuations are not amplified, so they may be
discarded to a good approximation in this analysis. The Green's
function $G$ satisfies
\begin{equation}
  \frac{1}{(1+\deltat)^2}\KG G(x_1,x_2) = \delta_4(x_1-x_2)
\end{equation}
with $\delta_4$ the delta function in $\measure{s}_4^2$.
As before we seek to solve $G$ as the Fourier transform
$G(\kb,\tau,\tau_2)$ of a perturbation expansion $G = G_0 + \delta
G $ in $\delta H$.  We change to conformal time $\tau$ and write
$G_0 \rightarrow a_b G_0$. The $O(1)$ equation is $\bop{3/2}G_0 =
\delta(t - t_2) / a_0$. (When writing $a_b$, we always mean the
unperturbed $a_b$ which satisfies $a_b = \e{H_0 t} =
-(H_0 \tau)^{-1}$, with $\tau$ the unperturbed conformal time. Variations in
$a_b$ are explicitly written as $\delta a_b$.)
This is the same as Eq.~\eqref{eq:u-unperturb} for the unperturbed four-%
dimensional Green's function, and so shares the
same solution.

The $O(\delta H)$ equation
differs from the four-dimensional case Eq.~\eqref{eq:perturb-fourd}.  It is
\begin{eqnarray}
  \nonumber
  \bop{3/2} \delta G & = & \left[ 2 k^2 G \frac{\delta a_b}{a_b} -
  \left( 3 \delta H + 2 \dot{\deltat} \right) \left( a_b
  \pd{G_0}{\tau} - a^2 H_0 G_0 \right) \right] \\
  \nonumber
  & & - \left(3 \frac{\delta a_b}{a_b} + 2 \deltat \right)
  \frac{\delta(\tau - \tau_2)}{a_b} \\
  & = &
  \label{eq:five-d-perturb}
  \tilde{\source}(\tau,\tau_2) - \left( 3 \frac{\delta a_b}{a_b} + 2
  \deltat \right) \frac{\delta(\tau - \tau_2)}{a_b} .
\end{eqnarray}
Nonetheless, this expression has exactly the same structure as
the four-dimensional problem: the operator on the left-hand side is
$\bop{3/2}$, which permits a solution for $\delta G$ as a transform
in the eigenfunctions
of $\bop{3/2}$.  The right-hand side takes the form of some source
term $\tilde{\source}$
and a contribution proportional to $\delta(\tau - \tau_2)$.
The difference lies in the explicit form of $\tilde{\source}$ and the
coefficient of the $\delta$-distribution.

The solution of Eq.~\eqref{eq:five-d-perturb} therefore comes out
to be
 \ba
  \delta G &=& \int_{-\infty}^{\infty} \frac{\measure{m}}{-m^2}
  \phi_m(\tau) \int_{-\infty}^0 \measure{\eta} \;
  \tilde{\source}(\eta,\tau_2) \phi_m(\eta) \nn
  & & - \left(3 \frac{\delta a_b}{a_b} + 2\delta g \right)
  G_0(\tau,\tau_2) .
 \ea
The conditions for the $\bop{3/2}$ or Fourier--Bessel transform
to make sense are the same as the four-dimensional case; in particular,
$\delta\rho$ must vanish as $\eta \rightarrow -\infty$ and obey
an equation of state $p = w \rho$ with $w > -1/3$ as $\eta \rightarrow 0$.
There is very little loss of generality in assuming that $\delta\rho$ dies
away in the asymptotic future as well as the asymptotic past.

One now writes the full two-point function in a manner analogous to
Eq.~\eqref{eq:fourd-fulltwopoint} and takes the limit $x_1 \rightarrow
x_2$ to give a power spectrum.  This part of the argument involves
manipulations very similar to those leading to Eq.~%
\eqref{eq:perturb-fourd-spectrum}, so we do not write them out explicitly.
One then takes the large scale limit
$k \rightarrow 0$.
The source function $\tilde{\source}$ obeys
similar asymptotics to Eq.~\eqref{eq:source-asymptotics}: in particular,
terms arising from $\partial G_0/\partial t$ vanish, so the terms
involving $\delta H$ and $\dot{\deltat}$ disappear.  Therefore, one has
 \ba
  \label{eq:perturb-fived-spectrum}
  \powerspectrum{T,5} &=& 8 \fourgrav^2 F^2
  \left( \frac{H_0}{2\pi} \right)^2 \bigg( 1 - 3
  \frac{\delta a_b}{a_b} - 2 \deltat \nn
  & & - 2 \int_{-\infty}^{\infty}
  \frac{\measure{m}}{m^2} \frac{\hat{\phi}_m(\tau)}{\tau}
  \int_{-\infty}^0 \measure{\eta} \; \frac{\hat{\phi}_m(\eta)}{\eta}
  \frac{\delta a_b}{a_b} \bigg) .
 \ea
This is the power spectrum, as seen on the brane, of gravitational
waves excited during the perturbed inflationary epoch.

Eq.~\eqref{eq:perturb-fived-spectrum} has the important property,
alluded to in the introduction, that as $\lambda \rightarrow \infty$
it does not go over to the corresponding four-dimensional form
$\powerspectrum{T,4}$.  This is because the contribution
$\deltat = \delta H / H$ does not vanish as $\lambda \rightarrow
\infty$.
\section{Consistency relations on perturbed backgrounds}
\label{sec:pconsistency}
One does not expect that a relationship between observable
quantities of the form $\powerspectrum{T}
/ \powerspectrum{\perturbation} \propto n_T$ should hold for either
of these perturbed universes.  In any case,
this relation is only true to first
order in the slow-roll expansion and is known to receive corrections
of a different functional form at next-order \cite{lidsey-liddle}.
For this reason, it is clear that the consistency relation as we have
written it here, as direct proportionality
between the ratio of $\powerspectrum{T}$
and $\powerspectrum{\perturbation}$ and the tensor spectral index $n_T$,
is only an approximation to whatever exact connexion exists
between $\powerspectrum{T}$, $\powerspectrum{\perturbation}$,
$n_T$ and $n_{\perturbation}$.
In particular, the next-order result involves all four observable
quantities.
However one can meaningfully ask about the manner in which this
approximate relationship is broken by introducing a perturbation $\delta H$.

For the purposes of this section, it is useful to recast the results
Eq.~\eqref{eq:perturb-fourd-spectrum} and
Eq.~\eqref{eq:perturb-fived-spectrum}
for the power spectra $\powerspectrum{T,4}$ and $\powerspectrum{T,5}$
in a slightly different form, by making the prefactor involve the full
perturbed $H$ rather than its fixed background value $H_0$.  This
corresponds simply to including a contribution $-2 \delta H / H_0$ in the
$O(\delta H)$ terms.  In fact, for the argument we are about to make
the detailed form of these terms is not important so we will denote them
collectively just by $\gamma$.  Then, comparing Eq.~%
\eqref{eq:perturb-fourd-spectrum} and Eq.~\eqref{eq:perturb-fived-spectrum},
\begin{equation}
  \label{eq:simple-fourd-spectrum}
  \powerspectrum{T,4} = \left( \frac{H}{2\pi} \right)^2 (1 + \gamma)
\end{equation}
and
\begin{equation}
  \label{eq:simple-fived-spectrum}
  \powerspectrum{T,5} = \left( \frac{H}{2\pi} \right)^2 (1 + \gamma
  - 2 \deltat) .
\end{equation}

Let us assume that the scalar field $\phi$ dominates the energy
density of the universe.
The matter power spectrum still \cite{wands-malik} satisfies
$\powerspectrum{\perturbation} =
(H/\dot{\phi})^2 \powerspectrum{\phi}$ in the four-dimensional and
brane case equally, where $\powerspectrum{\phi}$ is given by
Eq.~\eqref{eq:perturb-fourd-spectrum}.  Here, we assume that $H$ includes
corrections
owing to the perturbation $\delta\rho$ in the density, which might be
sourced by some perturbation $\delta\phi$ in the scalar field.
However, we will continue to assume that the four- and
five-dimensional Hamilton--Jacobi equations Eq.~\eqref{eq:fourd-hj} and
Eq.~\eqref{eq:fived-hj} hold, with $H$ including contributions of
$O(\delta\rho)$.  The tensor power spectrum in the four-dimensional
case is given by an identical
expression to its unperturbed counterpart, namely
$\powerspectrum{T,4} = 8 \fourgrav^2 \powerspectrum{\phi}$.
In the brane world, one has instead
\begin{equation}
  \powerspectrum{T,5} = 8 \fourgrav^2 F^2
  \powerspectrum{\phi} \left( 1 - 2 \deltat
  \right) ,
\end{equation}
as can be seen by inspection of Eq.~\eqref{eq:simple-fourd-spectrum}
and Eq.~\eqref{eq:simple-fived-spectrum}. Then the ratio
$\powerspectrum{T}/\powerspectrum{\perturbation}$ takes the
following form:
 \ba
  \label{eq:perturbed-ratios}
  \frac{\powerspectrum{T,4}}{\powerspectrum{S,4}} &=& \frac{32}
  {\fourgrav^2} \frac{H^{\prime 2}}{H^2}, \nn
  \frac{\powerspectrum{T,5}}{\powerspectrum{S,5}} &=& \frac{32 \scale^2}
  {\fourgrav^2} F^2 \frac{H^{\prime 2}}{H^2}\frac{1}{H^2 +
  \scale^2} \left( 1 - 2 \deltat \right) .
 \ea
As noted above, we are assuming that $\dot{\phi}$ and $H$ are related
by perturbed versions of the Hamilton--Jacobi equations, and
$H$ should be expanded as $H = H_0 + \delta H$.  This procedure is
rather approximate, but one is already obliged to introduce
approximations into the calculation when differentiating with respect
to $\log k$, and controlling each of these estimates is not trivial.
As discussed above, neither of these ratios are now equal to their
respective tensor indices $n_T$.

Let us deal with the four dimensional case first.  In any event, we
need an expression connecting observable quantities in four dimensions
with which to compare any such connexion in five dimensions.
In fact,
it is no longer easy to write down any connexion between the four
principal observables $\powerspectrum{T,4}$,
$\powerspectrum{\perturbation,4}$, $n_{T,4}$ and $n_{\perturbation,4}$.
By differentiating Eq.~\eqref{eq:simple-fourd-spectrum} and the expression
$\powerspectrum{\perturbation,4} = (\dot{\phi}/H)^2 \powerspectrum{\phi}$
for $\powerspectrum{\perturbation,4}$ one can find the spectral indices.
They are,
\begin{equation}
  \label{eq:pfourd-tensor-index}
  n_{T,4} = -\frac{4}{\fourgrav^2} \left( \frac{H'}{H} \right)^2 +
  \deriv{\gamma}{\, \log k},
\end{equation}
and
\begin{equation}
  \label{eq:pfourd-scalar-index}
  n_{\perturbation,4} - 1 = - \frac{8}{\fourgrav^2} \left( \frac{H'}{H}
  \right)^2 + \frac{4}{\fourgrav^2} \frac{H''}{H} +
  \deriv{\gamma}{\, \log k} .
\end{equation}
All of these expressions are assumed to be taken at first order
in the slow-roll expansion, and first order in $\gamma = O(\delta H)$.
Where products of two or more slow-roll parameters appear, or a product
of a quantity at $O(\delta H)$ and a slow-roll parameter, they are
assumed to be discarded together with contributions of $O(\delta H^2)$.
To express $n_{T,4}$ in terms of the other observables, one must find some
way to eliminate $d \gamma / d \, \log k$.
This quantity looks like a contribution of order $\text{(slow roll)}^{1/2}
\times O(\delta H)$, so we do not discard it.  (Because of this,
one can regard our result as a kind of $1 \frac{1}{2}$-order expansion.)
Since $\gamma$ involves a particular Fourier--Bessel transform of the
quite general source function $\source$ (or $\tilde{\source}$),
which does not appear in any other quantity such as $H$,
it would appear that the only way of eliminating
it is to introduce $n_{\perturbation,4}$.  The terms involving
$H'/H$ can be rewritten using Eq.~\eqref{eq:perturbed-ratios}, but there
is no way of rewriting $H''/H$ using only the observables under
discussion.

To circumvent this difficulty requires the introduction of extra
observable parameters with which to broaden the range of quantities
we can express using them.  In the present case, it is most convenient
to introduce the running $r_{T,4}$ of the tensor spectral index
which was defined in Eq.~\eqref{eq:index-running}.  This satisfies
\begin{equation}
  \label{eq:pfourd-tensor-running}
  r_{T,4} = - 8 \frac{\powerspectrum{T,4}}{
  \powerspectrum{\zeta,4}} \left( \frac{H''}{H} - \frac{H^{\prime 2}}{H^2}
  \right) + \derivtwo{\gamma}{[\log k]^2} .
\end{equation}
The second derivative of $\gamma$ is of order an $O(\delta H)$ term
multiplied by a slow-roll parameter, so it can be discarded at the
order to which we are working.  Combining Eq.~%
\eqref{eq:pfourd-tensor-running} with Eqs. \eqref{eq:pfourd-tensor-index}--%
\eqref{eq:pfourd-scalar-index} allows us to write an expression for
$n_{T,4}$ involving only quantities which are in principle observable:
\begin{equation}
  \label{eq:pfourd-consistency}
  n_{T,4} \approx \frac{3}{16} \frac{\powerspectrum{T,4}}
  {\powerspectrum{\perturbation,4}} + (n_{\perturbation,4} - 1)
  + \frac{r_{T,4}}{2\fourgrav^2} \frac{\powerspectrum{\perturbation,4}}
  {\powerspectrum{T,4}} .
\end{equation}
We write an approximate equality $\approx$ to indicate that this
result is true only to first order in a combined slow-roll/$O(\delta H)$
expansion.

The principal result of this paper is that no such comparable
relation between observational quantities exists in the brane world.
One can again obtain spectral indices using the usual argument; for example,
the tensor spectral index satisfies
\begin{equation}
  \label{eq:pfived-tensor-index}
  n_{T,5} = - \frac{4 \scale^2}{\fourgrav^2} F^2 \left( \frac{H'}{H} \right)
  ^2 \frac{1}{H^2 + \scale^2} + \deriv{\gamma}{\, \log k} -
  2 \deriv{\deltat}{\, \log k} .
\end{equation}
This involves an extra $O(\delta H)$ correction $d \deltat
/ d \, \log k$.  One can write similar expressions for
$n_{\perturbation,5}$, $r_{T,5}$ and so on which we shall suppress
because of their complexity.  However, none of these quantities
contains any counter-term with which one could balance the
$d \deltat / d \, \log k$ appearing in
Eq.~\eqref{eq:pfived-tensor-index}.  This is because no such contribution
occurs in the $\perturbation$ quantities, and higher derivatives
of $n_{T,5}$ will not contain $\delta H$ at first order: as
in four dimensions, second derivatives of $\gamma$ and $\delta H$
are at second order in the slow-roll/$O(\delta H)$ expansion.
Thus it is quite impossible to write any relationship between
observable quantities which respects Eq.~\eqref{eq:pfived-tensor-index}.

This is a rather stronger statement than that the four-dimensional
consistency relation no longer holds in the brane world: rather,
we have shown that there is no consistency relation at all, involving
$n_T$, that holds in this universe.  This means that all four
observables $n_{\perturbation}$, $n_T$, $\powerspectrum{\perturbation}$
and $\powerspectrum{T}$ as well as any running in them are
independent.  We consider this as a kind of manifestation of the well-known
result that the four-dimensional cosmological perturbation theory
does not constitute a closed system: one needs extra information about
the behaviour of the bulk.  The change in the behaviour of gravitational
quantities is a result of an off-brane effect which cannot be
reconstructed in terms of purely brane-based observables.

Although we have exhibited this property only in a single example model,
we expect such behaviour to be quite generic.  We would like to
stress that not only can one not write a consistency relation for the
brane world at finite $\lambda$, no such consistency relation exists
as $\lambda \rightarrow \infty$ either.  This is because the $\delta H/H$
contribution which gives trouble does not disappear in this limit;
it is a consequence of the fact that $\powerspectrum{T,5}$ does
not approach $\powerspectrum{T,4}$ as $\lambda \rightarrow \infty$.
\section{Conclusions}
\label{sec:conclude}
The brane universe offers the prospect of reproducing four-dimensional
physics from string vacua without the necessity of compactifying extra
dimensions on Planck--sized manifolds.  One would like to investigate
the possible consequences of scenarios of this type: apart from
signatures in the particle physics sector, the other major high energy
laboratory in which one could conceive of observing such consequences
is the early universe.  So it is important to study the implications
of early universe cosmology in the brane world scenario.

In this paper, we have developed a perturbation expansion for
the gravitational wave modes around the pure de Sitter case
$H = \text{constant}$.  This perturbation series can be used in the
brane world and in four dimensional equally.  We use this technology
to calculate the power spectrum of scalars and gravitational waves
as seen on the brane, or in four dimensions, and write a consistency
relation in the four-dimensional case.  We also show that no such
consistency relation exists in the brane world.

This analysis confronts a troubling feature of the brane world model:
it predicts an identical observational degeneracy in comparison with
the conventional four-dimensional cosmology.  We have shown, by
an explicit calculation, that degeneracies of this type are not
generic.  If one takes a four-dimensional cosmology exhibiting some
degeneracy, then one should not expect, in general, for this degeneracy
to remain when one the four-dimensional cosmology as a brane.
This result is important; a complete degeneracy would hinder any
attempt to observationally reconstruct the inflaton potential.

Our calculation relies on exploiting a technical device to calculate the
tensor power spectrum in a model perturbed around a de Sitter brane
carrying a single scalar field.  This extends the range of models in
which one knows how to solve for the spectrum of gravitational waves
produced during an inflationary epoch.  This is a hard problem, whose
complete solution is not yet understood (but see Ref. \cite{hawking-hertog}
for a calculation of the tensor spectrum in the case where the brane
carries a large $N$ CFT, using the AdS/CFT correspondence).
Our method will not easily
generalize to full case of arbitrary time evolution on the brane,
but may suggest future directions in which to proceed.  One such
possibility is to study the brane universe in explicitly static
Schwarzschild--Anti de Sitter (SAdS)
coordinates, where there is a holonomic timelike Killing vector
$\partial / \partial T$.  The graviton field equation is then
independent of $T$ and becomes an ordinary differential equation,
similar to the Regge--Wheeler equation of black hole perturbation
theory.  The brane appears as a Neumann boundary condition applied to
what is effectively a moving mirror, and it is possible that this framework
is accessible to analytic attack.  Our calculation does not yet include
back reaction from other fields on the brane, so it not general
enough (for example) to include other types of matter, or to generalize
to a second order result.

We have argued that one can associate the persistence of a consistency
relation with the property that $\powerspectrum{T,5}$ (and therefore
$n_{T,5}$) goes over to its four-dimensional counterpart as one
decouples the brane from the bulk by taking $\lambda \rightarrow \infty$.
This property holds for the exact de Sitter imbedding but need not hold
for more general cosmologies.  The central feature of this
correspondence is that the relationship between observables on the
brane is independent of the tension $\lambda$, and so is guaranteed
to match the four-dimensional result.  This property is enforced by
the differential equation \eqref{eq:odd-diff-eq}.
We exhibit directly a solution corresponding
to a marginally perturbed de Sitter geomety for which $\powerspectrum{T,5}$
does not approach its four-dimensional counterpart in the decoupling
limit, a consistency relation does not exist, and the relationship between
observables is not $\lambda$-independent.

One might worry that our analysis involves a small perturbation
propagating on top of an inflating universe.  By the usual inflationary
arguments this perturbation should decay exponentially quickly
as inflation proceeds, and become negligible.  Although this is true,
we conceive of the perturbation as a probe of gravity's
behaviour in the two cosmologies, regardless of its absolute magnitude.
Because we did not specify how the perturbation to the four-dimensional
cosmology carried by the brane was to be sourced, one can possibly imagine
it to derive from bulk effects, such as an impinging gravitational wave.
The comparison four-dimensional calculation should then be understood
as a formal device.  Alternatively, it is simple an imbedded cosmological
model in some open neighbourhood of the exact de Sitter solution.
However one interprets the calculation, however,
the underlying principle is the same: one does not recover four-dimensional
quantities after taking $\lambda$ to infinity.
\begin{acknowledgments}
DS is supported by a PPARC Studentship.  We would like to thank
David Wands for useful conversations regarding this work.
\end{acknowledgments}
\appendix
\section{The normalization function $F$}
\label{sec:diff-eq}
In this section we sketch how the normalization function $F$ and the
central differential equation Eq.~\eqref{eq:odd-diff-eq} are obtained.

One defines $F$ to satisfy $2 \mu F^2 = \efunction_0^2$, where
$\efunction_0$ is the zero-mode of $\Y$.  The $\efunction_\alpha$ are
normalized in the Sturm--Liouville measure arising from $\Y$, that is,
$2 \int \measure{y} \; n^2 \, \efunction_\alpha \efunction_\beta =
\delta_{\alpha\beta}$.  The factor of $2$ has been added to take
account of the other branch of the orbifold, since we work on $y \in
[0,y_h]$.  Because the $\efunction_0$ are independent of
$y$, this just says $2 \mu F^2 \int n^2 \; \measure{y} = 1$.

It is easy to evaluate this integral directly, but for the purposes of
obtaining Eq.~\eqref{eq:odd-diff-eq}, it is convenient to
employ the relation $n' = - \sqrt{H^2 + \scale^2 n^2}$ which arises
from the Einstein field equation.  In that case, the normalization
requirement depends only on the integral of the purely geometrical
quantity $n$,
\begin{equation}
  \label{eq:norm-require}
  2 \mu F^2 \int_0^1 \frac{n^2 \; \measure{n}}{\sqrt{H^2 + \ell^2
  n^2}} = 1 .
\end{equation}
This does not depend on a detailed knowledge of the form of $n$,
except through $n'$.  Here $\mu$ is the ratio $\fourgrav^2/\fivegrav^2$
of the four- and five-dimensional gravitational couplings and $\ell$
is the AdS radius, which in the case of vanishing four-dimensional
cosmological constant equals $\mu$.  This is the case throughout the
main body of this paper.
One now makes a trigonometric substitution to evaluate the integral.
The result is
\begin{equation}
  \label{eq:f-defn}
  \frac{\mu}{\ell} F^2 \left( \sqrt{1 + \frac{H^2}{\ell^2}} -
  \frac{H^2}{\ell^2} \arsinh \frac{\ell}{H} \right) = 1 .
\end{equation}
If $\mu = \ell$ then this result agrees with Refs. \cite{lmw, gorbunov}.
To derive Eq.~\eqref{eq:odd-diff-eq}, one can differentiate this
result directly, but it is easier to proceed as follows.  Multiply
Eq.~\eqref{eq:norm-require} by $H^2$ and differentiate logarithmically.
One finds,
\begin{equation}
  \label{eq:proto-diff-eq}
  \deriv{\, \log HF}{\, \log H} + \mu F^2 \deriv{}{\, \log H} \int_0^1
  \frac{n^2 \; \measure{n}}{\sqrt{H^2 + \scale^2 n^2}} = 1 .
\end{equation}
It is now easy to differentiate under the integral sign, and integrate
the resulting expression.  That gives
\begin{equation}
  \deriv{}{\, \log H} \int_0^1 \frac{n^2 \; \measure{n}}{\sqrt{H^2 +
  \scale^2 n^2}} = \frac{1}{\mu F^2} - \frac{1}{\sqrt{H^2 + \scale^2}}
  .
\end{equation}
Substituting this into Eq.~\eqref{eq:proto-diff-eq} gives the result
Eq.~\eqref{eq:odd-diff-eq}.  This relation was first noticed by the
authors of Ref. \cite{huey-lidsey-A}.

Because the normalization integral does not involve integrating over
a solution to the field equation $\Box \phi = 0$, one can interpret
this result as a statement about the dependence of the metric $g_{ab}$
on the initial conditions prescribed on the brane.
%
%

\end{document}